\documentclass[11pt]{iopart}

\usepackage{amsbsy,amssymb,mathrsfs}

\newcommand{\Ga}{\Gamma}
\newcommand{\al}{\alpha}
\newcommand{\be}{\beta}
\newcommand{\ga}{\gamma}
\newcommand{\de}{\delta}
\newcommand{\m}{\mu}
\newcommand{\n}{\nu}

\newcommand{\eps}{\epsilon}

\newcommand{\dphi}{\Ga}

\newcommand{\Si}{\Sigma}

\newcommand{\mc}{\mathcal}
\newcommand{\ce}{\mc{E}}

\newcommand{\cs}{N}
\newcommand{\cl}{\mc{L}}

\newcommand{\ca}{\mc{A}}

\newcommand{\ch}{\mc{H}}
\newcommand{\cq}{\mc{Q}}
\newcommand{\cg}{\mc{G}}
\newcommand{\cf}{\mc{F}}
\newcommand{\en}{n}
\newcommand{\nn}{\nonumber}
\newcommand{\ex}{\mc{C}}

\newcommand{\veh}{{\mbox{v}}}
\newcommand{\teh}{{\mbox{\tiny{T}}}}

\newcommand{\gee}{T}

\begin{document}

\title[1+1+2 gravitational perturbations on LRS class II space-times]{1+1+2 gravitational perturbations on LRS class II space-times: GEM vector harmonic amplitudes}

\author{R. B. Burston}

\address{Max Planck Institute for Solar System Research,
37191 Katlenburg-Lindau, Germany}
\eads{\mailto{burston@mps.mpg.de}}

\begin{abstract}
This is the second in a series of papers which considers first-order gauge-invariant and covariant gravitational perturbations to {\it locally rotationally symmetric} (LRS) class II space-times. This paper shows how to decouple a complex combination of the {\it gravito-electromagnetic} (GEM) 2-vectors with the 2-tensors describing the shear of the 2/3-sheets. An arbitrary harmonic expansion is then used along with an eigen-vector/value analysis of the first-order GEM system, analogous to the first paper in this series \cite{Burston2007GEMT}. This results in four real decoupled equations governing four real combinations of the harmonic amplitudes of the GEM 2-vectors and the (2/3-sheet) shear 2-tensors. Finally, these are categorized into polar and axial perturbations.
\end{abstract}

\pacs{04.25.Nx, 04.20.-q, 04.40.-b, 03.50.De, 04.20.Cv}
\maketitle

\section{Introduction}

The 1+1+2 gauge-invariant and covariant formalism \cite{Clarkson2003} is very well suited for describing gravitational perturbations to {\it locally rotationally symmetric} space-times \cite{Ellis1967,Stewart1968,Elst1996}. This was first presented in \cite{Clarkson2003} for an analysis of vacuum gravitational perturbations to a covariant Schwarzschild space-time and was followed by studies of electromagnetic perturbations to LRS class II space-times in \cite{Betschart2004,Burston2007EMVH} and LRS space-times in \cite{Burston2007EMBP}.  The first-order equations governing gravitational perturbations to LRS class II space-times were then presented in \cite{Burston2007GEMT} and for LRS space-times in \cite{Clarkson2007}.

The first paper in this series, hereafter Paper I, showed that the first-order {\it gravito-electromagnetic} (GEM) could be expressed in a 1+1+2 complex form which is conducive to decoupling \cite{Burston2007GEMT}.  Paper I showed how to derive a gauge-invariant and covariant second-order differential equation governing the complex GEM 2-tensor thus clearly demonstrating how it decouples from the remaining first-order 1+1+2 quantities. We further used an arbitrary tensor harmonic expansion and ultimately revealed that only particular combinations of the GEM 2-tensor harmonic amplitudes decouple and they were further categorized into polar and axial perturbations. 

In section \ref{backfdds}, we present  a brief review of the background space-time and the first-order perturbation variables. The reader is referred to Paper I and \cite{Clarkson2003} for a more comprehensive description. In Section \ref{dsacomd} we summarize two primary results from Paper I; the 1+1+2 GEM system may be expressed in a new complex form and there are four specific combinations of the GEM 2-tensor harmonic amplitudes which decouple. Then in Section \ref{dfsvcxZ} we choose new dependent variables and subsequently show how to decouple one of these. This is followed by a harmonic expansion ultimately revealing another four decoupled quantities.

We follow the notations and conventions of Paper I and \cite{Clarkson2003}.

\section{The background LRS class II space-time and the first-order perturbations}\label{backfdds}

 The background comprises the most general vacuum LRS class II space-time and is defined by six non-vanishing scalars\footnote{For further details on how these quantities arise from the decomposition of the standard 1+3 quantities into 1+1+2 form, see \cite{Clarkson2003}.}
\begin{eqnarray}
\mbox{LRS class II}:\{\ca, \phi,\Si,\theta,\ce, \Lambda\}.
\end{eqnarray}
The background Ricci identities for both $u^\m$ and $\en^\m$ and the Bianchi identities yields a set of evolution and propagation equations governing these scalars. These were first presented in \cite{Clarkson2003} for a covariant Schwarzschild space-time, generalized to non-vacuum LRS class II space-times in \cite{Betschart2004} and they were also reproduced in Paper I for the vacuum case. Here, $\ca $ is the radial acceleration of the four-velocity, $\theta$ and $\phi$ are respectively the expansions of the 3-sheets and 2-sheets and $\Si$ is the radial part of the shear of the 3-sheet. The radial part of the gravito-electric  tensor is  $\ce$,  and finally, $\Lambda$ is the cosmological constant.

The gravitational and energy-momentum perturbations on the background LRS class II space-times are quantities of first-order ($\epsilon$),
\begin{eqnarray}
\fl \mbox{first-order scalars:}    \hspace{-.2cm}  &\{\ch,\xi,\Omega,\mu, p, \cq,\Pi\}= \mc{O}(\eps),&\label{fqdas1}\\
\fl \mbox{first-order 2-vectors:}\hspace{-.2cm} &\{a^\m,\al^\m,\Omega^\m,\ca^\m,\Si^\m,\ce^\m,\ch^\m,V^\m,W^\m,X^\m,Y^\m,Z^\m,\cq^\m , \Pi^{\m}\}= \mc{O}(\eps),&\label{fqdas2}\\
\fl \mbox{first-order 2-tensors:}\,\,\,  \hspace{-.2cm} &\{\Si_{\m\n},\zeta_{\m\n},\ce_{\m\n},\ch_{\m\n}, \Pi_{\m\n} \}= \mc{O}(\eps),&\label{fqdas3}
\end{eqnarray}
as also defined in Paper I and \cite{Clarkson2003}. These first-order quantities given in \eref{fqdas1}-\eref{fqdas3} are all gauge-invariant under infinitesimal coordinate transformations, or more formally due to the Sachs-Stewart-Walker Lemma \cite{Sachs1964,Stewart1974}, as their corresponding background terms vanish. 

The first-order scalars can be described as follows, $\xi$ is the twisting of the 2-sheet, $\Omega$ is the radial part of the vorticity of the 3-sheet and the radial part of the gravito-magnetic tensor is $\ch$. The energy-momentum quantities, mass-energy density, pressure, radial heat flux and radial anisotropic stress are denoted respectively $\m$, $p$, $\cq$ and $\Pi$. 
The first-order GEM 2-tensors are $\ce_{\m\n}$ and $\ch_{\m\n}$, the 2-tensors describing the shear of the 2/3-sheets are respectively $\zeta_{\m\n}$ and $\Si_{\m\n}$ and finally, $\Pi_{\m\n}$ is the anisotropic stress which has been projected onto the 2-sheets.

Furthermore, there is also the issue of choosing a particular frame in the perturbed space-time (i.e. choosing the first-order four-velocity and radial vector) as also discussed in \cite{Clarkson2003}. In general, the first-order gauge-invariant 1+1+2 quantities will not be frame invariant as they naturally depend on this choice since their underlying definitions are typically just projections and contractions with the four-velocity and radial vector.

\section{The first-order complex 1+1+2 GEM system}\label{dsacomd}

Paper I showed that the first-order 1+1+2 GEM system may be expressed in new complex form according to 
\begin{eqnarray}
\fl \Bigl(\cl_\en+\frac 32\phi\Bigr) \ex_\m  + \de_\m \de^\al \Phi_\al +\frac 32\ce\Bigl[Y_\m -\phi a_\m-2\,\Bigl(\Si-\frac 23\theta\Bigr){\eps_\m}^\al \Omega_\al+\rmi2\de_\m \Omega \Bigr]  = \de_\m \cg \label{lnce} ,\\
\fl \Bigl(\cl_u-\frac 32\,\Si+\theta\Bigr) \ex_{\bar\m}+\rmi\,\de_\m (\eps^{\al\be} \de_\al \Phi_\be)\nn\\
  -\frac32\,\ce\, \left[\ca_\m\,\Bigl(\Si-\frac 23\,\theta\Bigr)+\phi\,(\Si_\m-{\eps_\m}^\al \Omega_\al+\al_\m) +W_\m-\rmi\,2\, \de_\m \xi \right]=\de_\m\cf, \label{luexm}\\
\fl(\cl_\en+\phi)\,\Phi_{\bar\m}+\de^\al \,\Phi_{\m\al}  -\,\frac12\,\de(\de_\m\, \Phi)-\rmi\,\frac32 \, \Sigma\, {\epsilon_\m}^\al\Phi_\al\,+\frac 32 \,\ce\,\Lambda_\m=\cg_\m,\label{lenphim}\\
\fl\Bigl(\cl_u-\Si+\frac 23\, \theta\Bigr) \Phi_{\bar\m}+ \rmi\,{\epsilon_\m}^\al \de^\be \Phi_{\al\be}+\rmi\, \frac 12\,{\epsilon_\m}^\al\,\left[ \ex_\al-(2\,\ca-\phi)\,\Phi_\al  \right]+\frac 32 \,\ce\,\Upsilon_\m=\cf_\m, \label{luphimdef}\\
\fl\Bigl(\cl_u +\frac52\Si+\frac13\theta\Bigr)\Phi_{\bar \m\bar\n}-\rmi{\epsilon_{(\m}}^\al \Bigl(\cl_\en +2\ca-\frac 12\, \phi\Bigr)\Phi_{\n)\al}  +\rmi\,{\epsilon_{\{\m}}^\al \de_{|\al|}\Phi_{\n\}}+\frac 32 \,\ce\,\Lambda_{\m\n} = \cf_{\m\n} \label{lnphimn},
\end{eqnarray} 
where  
\begin{eqnarray}
\fl \ex_\m := X_\m +\rmi \, \de_\m\ch, \qquad\Phi_\m := \ce_\m +\rmi \,\ch_\m \qquad\mbox{and} \qquad \Phi_{\m\n}:= \ce_{\m\n} + \rmi \, \ch_{\m\n}, \label{defpohids}
\end{eqnarray}
and further definitions were conveniently given by
\begin{eqnarray}
\fl\Upsilon_\m := \al_\m +\rmi\,{\epsilon_\m}^\al \ca_\al,\,\,\,\,\,
\Lambda_\m  :=  a_\m  +\rmi\,{\epsilon_\m}^\al(  \Si_\al+ {\epsilon_\al}^\be\Omega_\be ) \,\,\,\,\,\mbox{and}\,\,\,\,\,\,
\Lambda_{\m\n} := \Sigma_{\m\n}+\rmi \,{\epsilon_{(\m}}^\al \, \zeta_{\n)\al}.\label{newcomasd}
\end{eqnarray}
Furthermore, $\de_\m$ is the covariant 2-derivative associated with the 2-sheets, $\eps_{\m\n}$ is the Levi-Civita 2-tensor, $\rmi$ is the complex number and finally the energy-momentum sources are those defined in Paper I. We were then able to use the GEM system \eref{lnce}-\eref{lnphimn}, which consists of first-order derivatives only, to show the decoupling of the complex GEM 2-tensor, $\Phi_{\m\n}$. This was achieved by constructing an equation with higher order derivatives. Furthermore, a tensor harmonic expansion was used to  show that the specific combinations of the GEM 2-tensor harmonic amplitudes that decouple are,
\begin{eqnarray}
\mbox{Decoupled polar perturbations:     }\{\ce_\teh + \bar\ch_\teh,\ce_\teh- \bar\ch_\teh\} ,\nn\\
\mbox{Decoupled axial perturbations:     }\{\ch_\teh  + \bar\ce_\teh,\ch_\teh- \bar\ce_\teh\} .
\end{eqnarray}

\section{New dependent GEM variables}

In two recent papers, we successfully showed how to fully decouple EM perturbations to LRS class II space-times in \cite{Burston2007EMVH} and to LRS space-times in \cite{Burston2007EMBP}. Amongst the reasons that full decoupling could be achieved in these cases were, we expressed the 1+1+2 EM system in a complex form and furthermore, the system contained the first-order EM fields only. This is in contrast to the 1+1+2 complex GEM system \eref{lnce}-\eref{lnphimn} here as it does not purely depend on the 1+1+2 GEM quantities. By inspection, it can be seen to be coupled back to the 1+1+2 Ricci identities (see Paper I and \cite{Clarkson2007}) through terms such as $\Omega$, $\xi$, $\Lambda_\m$, $\Upsilon_\m$ and $\Lambda_{\m\n}$. These additional complications compared to the EM system did not seem to hinder the process of finding a decoupled equation for $\Phi_{\m\n}$ as the 1+1+2 Ricci/Bianchi identities were used in Paper I to remove these terms.   However, if one attempts to construct a decoupled equation for $\Phi_\m$ by taking the Lie derivative with respect to $u^\m$ of \eref{luphimdef}, then it follows that you must have an evolution equation for $\Upsilon_\m$ which then implies you need an evolution equation for the first-order quantity $\ca_\m$, for which there is none (again see the 1+1+2 Ricci identities presented in Paper I). Therefore, we seek to choose new dependent variables and construct higher-order derivatives to incorporate these miscellaneous terms. 

We begin by operating on \eref{lnce} and \eref{luexm} with $(\de^2+K)$ and using a linear combination of the results to result in \eref{eqdxi}. Similarly, by operating on both  \eref{lenphim}-\eref{luphimdef} with $(\de^2+K)$ yields \eref{lenpsi} and \eref{lupsi}, and finally by taking the 2-divergence of  \eref{lnphimn} results in \eref{divphi},
\begin{eqnarray}
\fl \Bigl(\cl_u-\frac 52\,\Si+\frac 53\,\theta\Bigr) \Xi_\m +\rmi\,{\epsilon_\m}^\al \Big(\cl_\en +\frac 52\,\phi\Bigr) \Xi_\al +\rmi\,  {\eps_\m}^\al(\de^2-K-3\,\ce)  \Psi_\al  \nn\\
-3\,\ce\,\Bigl[\Bigl(\Si-\frac 23\,\theta\Bigr)\dphi_\m+\rmi\,\phi\,{\eps_\m}^\al \dphi_\al \Bigr]  = \gee_\m \label{eqdxi},\\
\fl (\cl_\en+2\,\phi)\Psi_{\bar\m} -\rmi\,\frac 32\,\Si\,{\epsilon_\m}^\al \, \Psi_\al -\frac 12\,\Xi_\m+ (\de^2+K+3\,\ce) \dphi_\m = \tilde \cg_\m,\label{lenpsi}\\
\fl\Bigl(\cl_u-2\Si+\frac 43\theta\Bigr) \Psi_\m -\rmi {\eps_\m}^\al\Bigl(\ca-\frac 12\phi\Bigr)\Psi_\al +\rmi\frac 12{\eps_\m}^\al \Xi_\al  + \rmi(\de^2+K+3\ce) {\eps_\m}^\al \dphi_\al = \tilde\cf_\m, \label{lupsi}\\
\fl \Bigl(\cl_u +\frac32\,\Si+\theta\Bigr)\dphi_{\bar\m}
-\rmi\,{\epsilon_\m}^\al \Bigl(\cl_\en+2\,\ca +\frac 12\,\phi\Bigr) \dphi_\al+ \rmi\,\frac 12\,{\epsilon_\m}^\al \,\Psi_\al=\de^\al\cf_{\m\al}. \label{divphi}
\end{eqnarray}
where the new dependent variables have been defined
\begin{eqnarray}
\Xi_\m : = (\de^2+K)\ex_\m -3\,\ce\,\Bigl[ \phi\, \de^\al \zeta_{\m\al} +\Bigl(\Si-\frac 23\,\theta\Bigr) \,\de^\al \Si_{\m\al} +\de ^\al \Pi_{\m\al} \Bigr], \label{ximdef}\\
\Psi_\m:=(\de^2+K) \Phi_\m +\rmi\, 3\,\ce\, {\eps_\m}^\al \de^\be \Lambda_{\al\be},\label{psidefd}\\
\Ga_\m :=\de^\al \Phi_{\m\al},
\end{eqnarray}
and the energy-momentum sources
\begin{eqnarray}
\check \cf_\m := (\de^2+K) \cf_\m,\\
\check \cg_\m := (\de^2+K) \cg_\m,\\
\gee_\m := (\de^2+K) (\de_\m \cf +\rmi \,{\eps_\m}^\al \de_\al \cg ) + 6\,\ce\, \de^\al \cf_{\m\al}.
\end{eqnarray}

Thus, the new GEM system \eref{eqdxi}-\eref{divphi}  now involves only three quantities $\Xi_\m$, $\Psi_\m$ and $\Ga_\m$ and we will show in this paper how to decouple $\Psi_\m$,  and in a third paper how to decouple $\Xi_\m$. 

The process of constructing higher-order derivatives, as used here, to decouple particular quantities is traditionally standard practice. It is a ``textbook" example for EM in Minkowski space-time to decouple the electric and magnetic fields by constructing second-order differential equations from the first-order differential equations. Furthermore, these decoupled second-order equations have been extensively generalized for LRS class II space-times in \cite{Betschart2004,Burston2007EMVH} and for LRS space-times in \cite{Burston2007EMBP}. Since the GEM system \eref{eqdxi}-\eref{divphi} is substantially more complicated here, we have had to construct even higher order derivatives by operating with the 2-Laplacian. Also, the 2-Laplacian is a very well behaved operator and is comfortably manipulated using harmonic expansions in the coming sections. Furthermore, the underlying notion of modifying the Bianchi identities in this way has been explored in \cite{Fernandes1996} using the Newman-Penrose formalism.

As a final note, the 2-divergence of anisotropic stress 2-tensor ($\de^\al \Pi_{\m\al}$) is combined with the new definition of $\Xi_\m$ in \eref{ximdef}. We consider $\Pi_{\m\n}$ to be a known energy-momentum source and thus would usually place this term on the right-hand-side of the equations to indicate this. However, the way in which it arises in the definition for $\Xi_\m$ is very natural and so we choose leave it on the left-hand-side, but keep in mind that this is a known source.

\section{Decoupling $\Psi_\m$ and its vector harmonic amplitudes}\label{dfsvcxZ}

It is now possible to derive a decoupled equation governing $\Psi_\m$ by taking higher order derivatives again. Begin by taking the Lie derivative of \eref{lupsi} and the process is similar to that described in Paper I for the decoupling of $\Phi_{\m\n}$. However, one further step is required by operating once again with  $(\de^2+K+3\,\ce)$ to obtain the new result
\begin{eqnarray}
\fl\left\{\de^2+K+3\,\ce\right\} \Bigl\{[(\cl_u-3\,\Si+3\,\theta) \cl_u -(\cl_\en+\ca+4\,\phi)\cl_\en -V]\,\Psi_{\bar \m} \nn\\
 \qquad\qquad\qquad\qquad+\rmi\, {\epsilon_\m}^\al [-(2\,\ca+\phi)  \cl_u  +3\,\Si \,\cl_\en-U]\Psi_\al \Bigr\}\nn\\
\fl -\frac32\,\ce \,\Bigl[\phi\, {\cs_\m}^\al +\rmi\,\Bigl(\Si-\frac 23\,\theta\Bigr) {\eps_\m}^\al\Bigr] \nn\\
\times \Bigl[\rmi\,{\epsilon_\al}^\be \Bigl(\cl_u-\frac 72\,\Si+\frac 43\,\theta\Bigr) \Psi_\be +\Bigl(\cl_\en-\ca+\frac52\,\phi\Bigr)\Psi_{\al}\Bigr]  =S_\m \label{decoupsim},
\end{eqnarray}
where the terms related to the potentials are
\begin{eqnarray}
V := \de^2-\ca^2+\frac {11}4\,\phi^2+\ca\,\phi-4\,\ce -\frac 34\,\Si^2 +4\,\Si\,\theta -\frac 43\,\theta^2,\\
U:= \cl_u \ca-\frac 32\,\cl_\en\Si -4\,\ca\,(\Si-\frac 23\,\theta)-\phi\,(\frac {19}4\,\Si +\frac 43\,\theta),
\end{eqnarray}
and the energy-momentum source is
\begin{eqnarray}
\fl S_\m := \left\{\de^2+K+3\,\ce\right\}\Bigl\{ -\rmi\,2\,{\epsilon_\m}^\al(\de^2+K+3\,\ce) \de^\be \cf_{\al\be}\nn\\
\fl  \Bigl[{\cs_\m}^\al \bigl(\cl_u+\frac 12\,\Si+\frac 53\,\theta\Bigr)-\rmi\,{\eps_\m}^\al \Bigl(\cl_\en +2\,\ca+\frac 32\,\phi\Bigr)\Bigr] \Bigl[(\de^2+ K)(\cg_\al+\rmi\, {\epsilon_\al}^\ga \cf_\ga)\Bigr] \Bigr\} \nn\\
-\frac32\,\ce \,\Bigl[\phi\, {\cs_\m}^\al +\rmi\,\Bigl(\Si-\frac 23\,\theta\Bigr) {\eps_\m}^\al\Bigr]\,(\de^2+K) (\cg_\al+\rmi\,{\epsilon_\al}^\be \cf_\be).
\end{eqnarray}
Thus, \eref{decoupsim} clearly demonstrates the decoupling of $\Psi_\m$ and whilst this appears rather complicated, it is important to observe that this is gauge-invariant and covariant with a full description of energy-momentum sources. Furthermore, it is clear that \eref{decoupsim} will become more manageable once harmonic expansions have been made.

\subsection{Harmonic expansions}

In order to further decouple \eref{decoupsim}, we use an arbitrary 2-vector harmonic expansion as described in Paper I \cite{Betschart2004} according to
\begin{eqnarray}
\Psi_\m = \Psi_\veh\, Q_\m + \bar\Psi_\veh\,\bar Q_\m\qquad \mbox{and}\qquad S_\m = S_\veh\, Q_\m + \bar S_\veh\,\bar Q_\m,
\end{eqnarray}
and  \eref{decoupsim} becomes two complex equations of the form
\begin{eqnarray}
\fl f\, \Bigl\{\Bigl[\cl_u-4\,\Si+\frac {11}3\,\theta-\frac 3{2\, f} \,\ce\,\Bigl(\Si-\frac 23\,\theta\Bigr)\Bigr]\cl_u    -\Bigl[\cl_\en+5\,\phi+\ca+\frac3{2\,f} \,\ce\,\phi\Bigr]\cl_\en  - \check V\Bigr\}\,\Psi_\veh\nn \\
\fl +\rmi \,f\, \Bigl\{\Bigl[2\,\ca+\phi+\frac 3{2\,f} \,\ce\,\phi\Bigr] \cl_u  -\Bigl[3\,\Si-\frac 3{2\,f} \,\ce\,\Bigl(\Si-\frac23\,\theta\Bigr) \Bigr]  \cl_\en +\check U_V\Bigr\} \bar \Psi_\veh = S_\veh \label{psiv},
\end{eqnarray}
\begin{eqnarray}
\fl f\,\Bigl\{ \Bigl[\cl_u-4\,\Si+\frac {11}3\,\theta-\frac 3{2\, f} \,\ce\,\Bigl(\Si-\frac 23\,\theta\Bigr)\Bigr]\cl_u   -\Bigl[\cl_\en+5\,\phi+\ca+\frac3{2\,f} \,\ce\,\phi\Bigr]\cl_\en  - \check V\Bigr\}\,\bar\Psi_\veh\nn \\
\fl -\rmi \,f\, \Bigl\{\Bigl[2\,\ca+\phi+\frac 3{2\,f} \,\ce\,\phi\Bigr] \cl_u  -\Bigl[3\,\Si-\frac 3{2\,f} \,\ce\,\Bigl(\Si-\frac23\,\theta\Bigr) \Bigr]  \cl_\en +\check U\Bigr\}  \Psi_\veh = \bar S_\veh\label{barpsiv}
\end{eqnarray}
where 
\begin{eqnarray}
\fl f := -\frac{k^2}{r^2} +5\,K +3\,\ce,\\
\fl \check V := f-\ca^2+\frac {19}4\,\phi^2+\ca\,\phi -5\,\ce-\frac 74\,\Si^2 +\frac {19}3 \,\theta\,\Si -\frac{22}{9}\,\theta^2 ,\nn\\
+\frac3{2\,f }\,\ce\, \Bigl[\phi\,(3\,\phi-\ca)-\Bigl(\Si-\frac 23\,\theta\Bigr)\Bigl(4\,\Si-\frac53\,\theta\Bigr)\Bigr] ,\nn\\
\fl \check U := \cl_u \ca-\frac 32\,\cl_\en \Si -5\,\ca\,\Bigl(\Si-\frac 23\,\theta\Bigr) -\phi\Bigl(\frac{27}4\,\Si+\theta\Bigr) \nn\\
-\frac3{2\, f} \,\ce\, \Bigl[ \ca\,\Bigl(\Si-\frac 23\,\theta\Bigr) +\phi\Bigl(\Si+\frac 13\,\theta\Bigr)\Bigr].
\end{eqnarray}
and $k^2$ and the scalar $r$ arise from the harmonic expansions of the 2-Laplacian.

Similar to the analysis of the tensor harmonics in Paper I, \eref{psiv}-\eref{barpsiv} are invariant under the simultaneous transformation of the form $\Psi_\veh\rightarrow \bar\Psi_\veh$ and $\bar\Psi_\veh \rightarrow \Psi_\veh$ (when the energy-momentum sources vanish, i.e.  $S_V=\bar S_V =0$). Therefore, the eigen-vector/value analysis we presented in \cite{Burston2007EMBP} also applies here and this shows that in order to decouple this system one needs to construct the complex combinations
\begin{eqnarray}
\Psi_\pm := \Psi_\veh \pm \rmi\,\bar\Psi_\veh \qquad\mbox{and}\qquad S_\pm := (S_\veh \pm\rmi\, \bar S_\veh), \label{psiupmdef}
\end{eqnarray}
 and we get two decoupled complex equations of the form
\begin{eqnarray}
\fl  f\,\Bigl\{\Bigl[\cl_u-4\,\Si+\frac {11}3\,\theta+(2\,\ca+\phi)-\frac {3\,\ce}{2\, f} \,\Bigl(\Si-\frac 23\,\theta-\phi\Bigr)\Bigr]\cl_u  \nn\\
\fl   -\Bigl[\cl_\en+5\,\phi+\ca+3\,\Si-\frac {3\,\ce}{2\,f} \,\Bigl(\Si-\frac23\,\theta-\phi\Bigr) \Bigr]\cl_\en  - \tilde V+ \tilde U\Bigr\}\,\Psi_+ =S_+ ,\label{eqnfppsip}
\end{eqnarray}
\begin{eqnarray}
\fl  f\,\Bigl\{\Bigl[\cl_u-4\,\Si+\frac {11}3\,\theta-(2\,\ca+\phi)-\frac {3\,\ce}{2\, f} \,\Bigl(\Si-\frac 23\,\theta+\phi\Bigr)\Bigr]\cl_u  \nn\\
\fl   -\Bigl[\cl_\en+5\,\phi+\ca-3\,\Si+\frac {3\ce}{2\,f} \,\Bigl(\Si-\frac23\,\theta +\phi \Bigr) \Bigr]\cl_\en  -\tilde  V-\tilde U\Bigr\}\,\Psi_- =S_- .\label{eqnfppsim}
\end{eqnarray}
where the $``\pm"$ are all relative.  

Since the differential operator in \eref{eqnfppsip}-\eref{eqnfppsim} is completely real and the dependent variables, $\Psi_\pm$, are complex, then there are in fact four real decoupled equations which can be readily found by taking the real and imaginary parts of \eref{eqnfppsip}-\eref{eqnfppsim} separately.

It is now of interest to see how these relate back to the original GEM 2-vectors and other 1+1+2 quantities. We expand the 2-tensors describing the shear of the 2/3-sheets using the arbitrary tensor harmonics defined in Paper I according to
\begin{eqnarray}
\Si_{\m\n} = \Si_T\, Q_{\m\n} +\bar \Si_T \,\bar Q_{\m\n} \nn \qquad \mbox{and}\qquad  \zeta_{\m\n} = \zeta_T \, Q_{\m\n} +\bar \zeta_T \,\bar Q_{\m\n}.
\end{eqnarray}
Then, by using the definitions \eref{newcomasd}, \eref{defpohids}, \eref{psidefd} and \eref{psiupmdef},  the four quantities that each satisfy a decouple equation are
\begin{eqnarray}
\Re[\Psi_+] =  h\,\Bigl[\Bigl( \ce_\veh+\frac 32\,\ce\, r\,\zeta_\teh\Bigr)-(\bar \ch_\veh+\frac 32\,\ce\, r\,\Si_\teh\Bigr)\Bigr] , \\
\Re[\Psi_-] =  h\,\Bigl[ \Bigl(\ce_\veh+\frac 32\,\ce\, r\,\zeta_\teh\Bigr)+(\bar \ch_\veh+\frac 32\,\ce\, r\,\Si_\teh\Bigr)\Bigr] ,\\
\Im[\Psi_+] = h \,\Bigl[ \Bigl(\ch_\veh+\frac 32\,\ce\,r\,\bar \Si_\teh\Bigr) +\Bigl( \bar\ce_\veh -\frac 32\,\ce\,r\,\bar \zeta_\teh\Bigr)\Bigr] ,\\
\Im[\Psi_-] = h \,\Bigl[ \Bigl(\ch_\veh +\frac 32\,\ce\,r\,\bar \Si_\teh\Bigr)- \Bigl(\bar\ce_\veh -\frac 32\,\ce\,r\,\bar \zeta_\teh\Bigr)\Bigr] ,
\end{eqnarray}
where
\begin{eqnarray}
h := 2\, K-\frac{k^2}{r^2} .
\end{eqnarray} 
Therefore, the factor $h$ may be differentiated out becoming a common factor and then the four specific combinations which decouple are separated into their polar and axial parts according to
\begin{eqnarray}
\fl \mbox{Polar:}  \Bigl\{ \Bigl( \ce_\veh+\frac 32\,\ce\, r\,\zeta_\teh\Bigr)-\Bigl(\bar \ch_\veh+\frac 32\,\ce\, r\,\Si_\teh\Bigr), \Bigl(\ce_\veh+\frac 32\,\ce\, r\,\zeta_\teh\Bigr)+\Bigl(\bar \ch_\veh+\frac 32\,\ce\, r\,\Si_\teh\Bigr)\Bigr\},\\
\fl\mbox{Axial:}  \Bigl\{ \Bigl(\ch_\veh+\frac 32\,\ce\,r\,\bar \Si_\teh\Bigr) +\Bigl( \bar\ce_\veh -\frac 32\,\ce\,r\,\bar \zeta_\teh\Bigr) , \Bigl(\ch_\veh +\frac 32\,\ce\,r\,\bar \Si_\teh\Bigr)- \Bigl(\bar\ce_\veh -\frac 32\,\ce\,r\,\bar \zeta_\teh\Bigr) \Bigr\}.
\end{eqnarray}
Finally, it is also clear that if one were to integrate the four decoupled equations (a highly non-trivial task) then linear combinations the solutions would reveal each of  $\Bigl( \ce_\veh+\frac 32\,\ce\, r\,\zeta_\teh\Bigr)$, $\Bigl(\bar \ch_\veh+\frac 32\,\ce\, r\,\Si_\teh\Bigr)$,  $\Bigl(\ch_\veh+\frac 32\,\ce\,r\,\bar \Si_\teh\Bigr)$ and $\Bigl(\bar\ce_\veh -\frac 32\,\ce\,r\,\bar \zeta_\teh\Bigr)$.

As a final comment, we note that one of these combinations is related to the significantly simpler analysis presented in \cite{Clarkson2003}. Therein, they analyzed a covariant Schwarzschild space-time for which the background consists of only three non-vanishing scalars $(\ca,\phi,\ce)$ and were able to show that $(\bar \ch_\veh+\frac 32\,\ce\, r\,\Si_\teh)$ satisfies a Zerilli equation. However, they are yet to show whether their result also holds true for arbitrary LRS class II space-times.

\section{Summary}

We have shown how to decouple particular combinations of the GEM 2-vectors and the 2-tensors defining the shear of the 2/3-sheets.  The gravito-electric 2-vector harmonic amplitudes are combined with the 2-tensor amplitudes of the shear of the 2-sheet, whereas the gravito-magnetic 2-vector amplitudes are combined with the amplitudes of the 2-tensor describing the shear of the 3-sheet.

\end{document}